\documentclass[preprint]{aastex62}
\usepackage{graphicx}
\DeclareGraphicsExtensions{.png,.jpg,.eps,.pdf}

\newcommand{\be}{\begin{eqnarray}}
\newcommand{\ee}{\end{eqnarray}}

\newcommand{\ehat}{ \hat{e}}

\newcommand{\ISOIS}{IS$\odot$IS~}

\usepackage{amsmath}
\usepackage{amsfonts}
\usepackage{amssymb}
\usepackage{lineno}
\usepackage{color}
\submitjournal{ApJ}
\accepted{December 17, 2020; December 31, 2020}

\begin{document} 

\newcommand{\revone}[1]{#1}
\newcommand{\revtwo}[1]{#1}
\newcommand{\revthree}[1]{#1}
\newcommand{\revfour}[1]{#1}
\newcommand{\remfour}[1]{}

\title{Switchbacks Explained: Super-Parker Fields -- the Other Side of the Sub-Parker Spiral}

\author{N. A. Schwadron}
\affiliation{University of New Hampshire, Durham, NH, 03824}
\affiliation{Department of Astrophysical Sciences, Princeton University, Princeton, NJ, 08544}

\author{D. J. McComas} 
\affiliation{Department of Astrophysical Sciences, Princeton University, Princeton, NJ, 08544}

\begin{abstract} 
We provide a simple geometric explanation for the source of switchbacks and associated large and one-sided transverse flows in the solar wind observed by Parker Solar
Probe.  The more radial, Sub-Parker Spiral structure of the heliospheric magnetic
field observed previously by Ulysses, ACE, and STEREO is created within rarefaction regions where footpoint motion from the source
of fast into slow wind at the Sun creates a magnetic field line connection across
solar wind speed shear.  Conversely, when footpoints move from the source of slow wind into faster wind, a Super-Parker Spiral field structure is formed: below the Alfv\'en critical point, one-sided transverse field-aligned flows develop; above the Alfv\'en critical point, the field structure contracts between adjacent solar wind flows, and
the radial field component decreases in magnitude with distance from the Sun,
eventually reversing into a switchback. The Sub-Parker and Super-Parker
Spirals behave functionally as opposites. Observations from Parker  Solar Probe
confirm the paucity of switchbacks within rarefaction regions and immediately outside
these rarefaction regions, we observe numerous switchbacks in the magnetic field that
are directly associated with abrupt transients in solar wind speed. 
The \remfour{magnetic field strength, the} radial component of the magnetic field, the speed gradients, 
\revfour{radial} Alfv\'en speed, and \remfour{plasma beta} \revfour{the ratio of the sound speed to the radial Alfv\'en speed} 
 all conform to predictions based on the Sub-Parker and
Super-Parker Spirals within rarefaction regions and solar wind speed enhancements
(spikes or jets), respectively.  Critically, the predictions associated with the Super-
Parker Spiral naturally explain the observations of switchbacks being associated with
unexpectedly large and one-sided tangential flows.
\end{abstract} 

\section{Introduction}
\label{sec:intro}

The  solar wind rapidly accelerates in the corona at $\sim 2$ -- 5 R$_s$ and then becomes super-Alfv\'enic at 
$\sim$10-20  R$_s$ 
\cite[]{Katsikas:2010,Goelzer:2014}. It is this latter transition where the solar wind ram pressure becomes dominant, 
overcoming both the magnetic and thermal pressure. The processes that transfer energy and 
dissipate this energy to  heat the corona and power the  solar wind remain critical  questions in heliophysics and astrophysics, 
and are at the heart of 
the scientific motivations for \emph{Parker Solar Probe} (PSP)  \cite[]{McComas:2007,Fox:2016}. 

On PSP, the solar wind is observed by the Solar Wind Electrons Alphas and Protons Investigation (SWEAP) \cite[]{Kasper:2016} and the magnetic field by the Electromagnetic Fields Investigation (FIELDS) \cite[]{Bale:2016}. The Integrated Science Investigation of the Sun (\ISOIS) instrument suite \cite[]{McComas:2016} provides comprehensive measurements of energetic particles  over the
range 0.02 -- 200 MeV/nucleon.  


 PSP  observes thousands of intervals (duration from seconds to tens of minutes) where the speed of the solar wind flow suddenly jumps 
 and includes a large, one-sided transverse flow, while simultaneously the magnetic field orientation rotates through large angles, before returning to roughly the prior solar wind conditions. The observed switchbacks (radial magnetic field reversals) are associated with the change in magnetic field direction and velocity spikes associated with the sharp increase in solar wind speed \citep{Bale:2019,Kasper:2019,deWit:2020,Horbury:2020,Mozer:2020,Rouillard:2020, Tenerani:2020}.

 Magnetic switchbacks have been studied extensively in fast wind from coronal holes \cite[e.g.,][]{Kahler:1996} at 1 au, and beyond 1 au with Ulysses \cite[e.g.,][]{Balogh:1999,Yamauchi:2004, Neugebauer:2013}.  Observations of switchbacks have also been made  inside 1 au with Helios \cite[]{Borovsky:2016,Horbury:2018} prior to the observations by  PSP.  Strong magnetic-field deviations from the Parker Spiral are observed where there are local increases in the radial solar wind speed \cite[]{Michel:1967}, and associated with one-sided or pulsed Alfv\'enic fluctuations \cite[]{Gosling:2009, Gosling:2011}.  The one-sided nature of switchbacks is especially clear in PSP observations.  \cite{Kasper:2019} states that ``Transients, including the Alfv\'enic jets, are one-sided, in that if the field rotates more than $\sim$60$^\circ$, then $B_T$ is always positive and $V_{pT}$ always exceeds 33 km s$^{-1}$.'' (Here, $B_T$ is the tangential magnetic field component, and $V_{pT}$ is the proton tangential component.) 
 
The transverse flows observed by PSP far exceed those in the Weber–Davis model \cite[]{Weber:1967} where the lower corona is taken to rotate rigidly at the mean rotational period of the Sun.  To this point \cite{Kasper:2019} state: "The large rotational velocities measured .. exceed the value in the axisymmetric Weber-Davis model, 
in which $V_{pT}(R_A) < 0.1\Omega_\odot R_A$, by more than an order of magnitude." Here, $\Omega_\odot$ refers to the solar rotation rate and $R_A$ refers to the Alfv\'en radius. For an Alfv\'en point of 15 $R_s$, \revthree{the $V_{pT}(R_A) < 3$ km s$^{-1}$. } Thus, the one-sided transverse flows are a key observable from PSP that any switchback theory must also explain.

Table 1 summarizes these key observations from PSP related to switchbacks and tangential flows. This paper provides a simple and natural geometric explanation for all of these apparently disparate observations that unifies the interpretation of switchbacks and the transverse flows observed by PSP. 

\begin{deluxetable}{ll}
\tablecaption{PSP observations of switchbacks and transverse flows \label{table:observations}}
\tablehead{
\colhead{Obs.} & \colhead{Reference} }
\startdata
Switchbacked magnetic field & \cite{Kasper:2019, Bale:2019}  \\
Transient jets or pulsations   & \cite{Kasper:2019}  \\
One-sided  tangential flows  & \cite{Kasper:2019} \\
Large transverse, co-rotational flows & \cite{Kasper:2019} \\
Alfv\'enic correlation $\delta \mathbf{v}$ to $\delta \mathbf{B}$ & \cite{Bale:2019} \\
Anti-correlated plasma density $n_e$ and $|\mathbf{B}|$ & \cite{Bale:2019} 
\enddata
\end{deluxetable}

There are already a variety of models and conjectures to explain various (although not all) aspects of switchbacks.  One example relates to non-linear shear driven turbulence \cite[]{Ruffolo:2020}.  In the presence of large speed gradients within the solar wind, there are a number of important effects that should be considered. As already indicated, the situation where faster wind outruns slower wind leads to the formation of rarefaction regions. However, if speed gradients exist across magnetic flux tubes,  the effects of non-linear shear driven turbulence can result in switchbacks  in the magnetic field \cite[]{Ruffolo:2020}.  Prior remote observations  \cite[]{DeForest:2016} show a transition from striated solar coronal structures to more isotropic  ``flocculated''  fluctuations in the transition just outside the Alfv\'en critical point.  This transition in the geometry of solar wind structures is powered by the relative velocities of adjacent coronal magnetic flux tubes.

Another conjecture relates to the effects of interchange reconnection (ICX) on the magnetic structure of the solar wind.  \cite{Fisk:2020} argue that the large transverse flows observed by PSP are a result of transverse flows in the corona and are part of a closed global circulation pattern of magnetic flux open to the solar wind. The circulation pattern at the Sun is sustained by the combined effects of differential motion and interchange reconnection \cite[]{Fisk:2001}.  \revtwo{\cite{Zank:2020} \revthree{develop} an evolution equation for the development of switchbacks resulting from interchange reconnection between coronal loops and the open magnetic field. Results from the model include complex aggregated groups of switchbacks  and an example event from the model resembles PSP observations. }

The concept of interchange reconnection was developed by \cite{Crooker:2002} to explain how magnetic flux injected by coronal mass ejections can be reduced without disconnecting magnetic fields entirely. 
Coronal mass ejections (CMEs) originate in closed magnetic field structures and add magnetic flux to the heliosphere as they move away from the Sun. Observations of ejecta in the solar wind show signatures including counter-streaming suprathermal electrons that show outflowing electron \revthree{heat flux moving in} both directions along closed field lines \cite[e.g.,][]{Gosling:1987}. These observations confirm the addition of magnetic flux from ejecta, and show that without changing the magnetic topology of CME ejecta, the total flux in the heliosphere would continue to increase, which leads to a ``magnetic field magnitude catastrophe'' \cite[]{Gosling:1975, McComas:1995a}. Disconnection was pictured as magnetic reconnection between magnetic field lines to create U-shaped structures that are disconnected entirely from the Sun and are then convected through the heliosphere by the solar wind \cite[]{McComas:1991}. While disconnection is one possible solution to the magnetic field catastrophe, it was noted that signatures of heat flux dropouts \cite[]{McComas:1989} should be associated with \revthree{disconnection. However,} the number of heat flux dropouts observed were found to be at least a factor of 4 too small to account for the loss of magnetic flux from that added by coronal mass ejections \cite[]{McComas:1992}. \cite{Crooker:2002}  suggested that the primary flux balancing mechanism is instead from interchange reconnection associated with magnetic reconnection between closed fields from coronal mass ejecta and open magnetic fields in the surrounding solar wind. \revthree{Interchange reconnection was considered as way for open magnetic fieldlines} to reconnect with loops in the corona, and thereby  enable large-scale redistribution of open magnetic fields beyond coronal hole boundaries \cite[]{Fisk:1999,Fisk:2001}. 

The conjecture that footpoint motion is driven by differential motion in coronal holes and interchange reconnection beyond coronal holes \revthree{was made previously by \cite{Fisk:1999} and \cite{Fisk:2001}. The basis of this conjecture is rooted in the difference between coronal holes
and the surrounding source regions of solar wind, which remains an area
of active research. Potential field models that approximate the corona as having zero current below the source surface
radius ($\sim$ 3 solar radii) show that open magnetic flux originates predominantly from
coronal holes \cite[]{Schatten:1969}.  Footpoint motions are driven by the Sun's differential motion and therefore 
convect open magnetic field lines continually across coronal hole boundaries \cite[]{Fisk:1999,Fisk:2001,Schwadron:2002c}. Outside coronal holes, interchange reconnection between open magnetic field lines and large loops 
may continue to facilitate the motion of the open magnetic field footpoints. 
The support for this view of footpoint motion is rooted in remote solar observations: 
\begin{itemize}
\item The photosphere rotates differentially \cite[e.g., ][]{Snodgrass:1983};
\item Coronal holes tend to rotate rigidly with the Sun, at approximately the equatorial rotation rate  \cite[e.g.,][]{Bird:1990}. 
\end{itemize} 
In this picture, coronal holes represent the 
boundary where there is a  transition between the open
magnetic flux concentrated within the coronal hole and the more distributed
open magnetic flux that moves through differential rotation and interchange reconnection beyond the coronal hole.}

\revthree{This picture of the Sun's open magnetic field has remained difficult to verify observationally. } One observational signature of footpoint motion was found in rarefaction regions where the magnetic field line is stretched between faster solar wind that streams out ahead of slower wind \cite[]{Murphy:2002,Schwadron:2002c}, \revthree{ leading to the formation of the Sub-Parker Spiral \cite[]{Schwadron:2004a,Schwadron:2005a}. Note that rarefaction regions are formed where faster solar wind outruns slower solar wind, and these structures are characterized by an almost monotonic decrease in the speed of solar wind. Rarefaction regions often map back to regions with small longitudinal extents on the Sun, termed ``dwells''  \cite{Schwenn:1990}. Because Stream interfaces typically co-rotate with the Sun, co-rotating rarefaction regions are formed from the trailing edge of coronal holes \cite[]{Smith:2000}. } 

Without footpoint motion, magnetic field lines are not connected between faster and slower wind, and conform to a Parker Spirals associated with their various solar wind speeds. In contrast, the straightening of the magnetic field in the rarefaction region due to footpoint motion at the Sun creates a magnetic field structure with a larger radial component than the Parker Spiral magnetic field.  The deviations
in field direction are extremely prominent and commonly observed by Ulysses in co-rotating rarefaction 
regions \citep{Murphy:2002,Schwadron:2004a,Schwadron:2005a}. The magnetic structure in rarefaction regions is referred to as the \emph{Sub-Parker Spiral} \cite[]{Schwadron:2004a,Schwadron:2005a} as it is less tightly wound than would be expected for the observed solar wind speed.  Another interpretation \cite[]{Gosling:2002} associated radial magnetic fields with abrupt temporal changes in the solar wind: the ``radially directed kink in the magnetic field connects the different spirals associated with the faster and slower flows immediately preceding and following the temporal flow speed discontinuity.'' This description is appropriate prior to or without the development of a rarefaction region.

In this paper, we show that the Sub-Parker Spiral depends on the direction of footpoint motion relative to the gradient between fast and slow solar wind. If the direction of footpoint motion is reversed with respect to the solar wind speed gradient, a different and distinct magnetic field structure is produced in which the field line connection between fast and slow wind contracts and ultimately reverses the magnetic field, producing a switchback. We begin by conceptually describing the  Super-Parker Spiral and its obvious extension to switchbacks as Super-Parker Spirals  in \S\ref{sec:concept}. In 
 \S\ref{sec:observations},  \revthree{we use observations from SWEAP and FIELDS to test for the presence of the Sub-Parker Spiral, and  the Super-Parker field structures. We report observations from several rarefaction regions and in nearby intervals of switchbacked fields and compare them to a simple quantitative model for such magnetic structures (see Appendix).  }
Finally, in \S\ref{sec:conk} we discuss the implications of this new interpretation and bring closure on the fundamental source for magnetic switchbacks and their uniquely one-sided transverse flows. 


\section{From Sub-Parker to Super-Parker Spirals}
\label{sec:concept}

\begin{figure}
\includegraphics[width=\columnwidth]{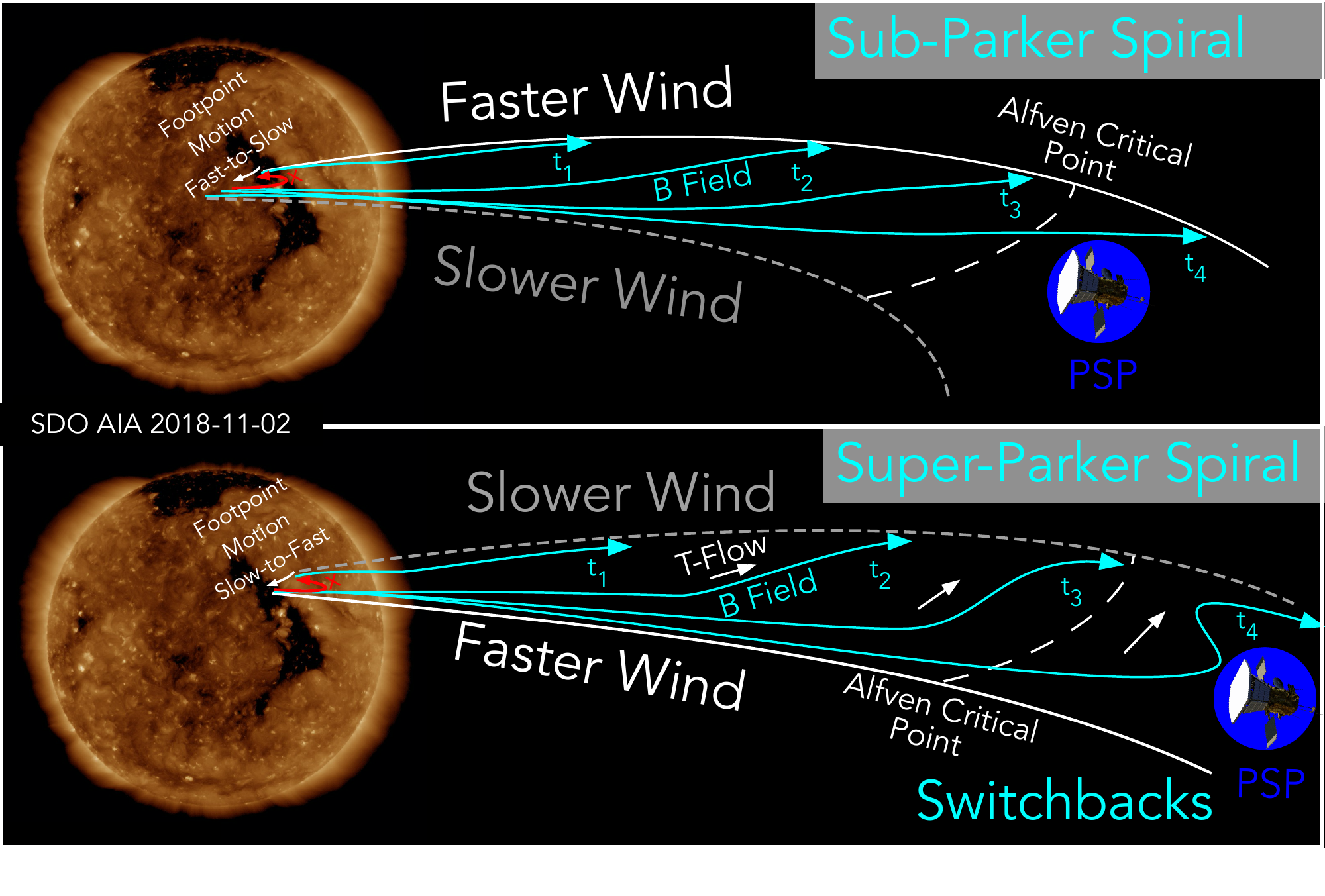} 
\caption{
The Sub-Parker Spiral (top panel) and the Super-Parker Spiral (bottom panel) result from footpoint motion between source regions of fast and slow solar wind. In the case of the Sub-Parker Spiral, footpoint motion from the source of fast to slow wind creates a fieldline connection that gets straightened as fast wind drags out magnetic fieldlines more quickly than the slow wind. The Sub-Parker Spiral is associated with magnetic fieldlines with larger radial components than the Parker Spiral. When footpoint motion is reversed (bottom panel), and footpoints move from the source of slow into fast wind, then the wind shear kinks the magnetic field. In this case, the faster wind moves along the magnetic field below the Alfv\'en critical point, forming compressions and tangential flows and develops into a switchback above the Alfv\'en critical point.
}
\label{fig:f1-switchback}
\end{figure}

The concept of the Sub-Parker Spiral is illustrated in Figure \ref{fig:f1-switchback} (top panel). Near the Sun, footpoint motion from the source of coronal-hole associated fast wind into slow wind creates a magnetic fieldline connection across the interface between fast and slow solar wind.  The fast solar wind draws the magnetic field out more quickly than the slow solar wind, therefore forming a rarefaction region. The magnetic field, stretched between faster solar wind flow and slower solar wind becomes increasingly radial with distance from the Sun.  The basic prediction in this case is the association between rarefaction and the Sub-Parker Spiral magnetic field, which has a stronger radial component than the Parker Spiral. \revthree{A recent study by \cite{Schwadron:2020} shows that the Sub-Parker Spiral provides relatively short fieldline connections from the PSP spacecraft to the compressions and shocks surrounding co-rotating interaction regions in the inner heliosphere (at $\sim$0.7 -- 10 au). These shorter fieldline connections are essential in explaining the persistence of energetic particles from $\sim 100$ keV to $>$ MeV observed by \ISOIS up to $\sim$1 week after the passage of a stream interface. }

The opposite configuration or Super-Parker Spiral is illustrated in the bottom panel of Figure \ref{fig:f1-switchback}. The fundamental difference between the Sub-Parker Spiral and the Super-Parker Spiral is the direction of footpoint motion. In the case of the Super-Parker Spiral, footpoints move from the source of slow wind into faster wind. Below the Alfv\'en critical point,  \revthree{the plasma flows along the magnetic field, and magnetic field tension resists the growth of fieldline curvature.} Therefore, faster wind overtakes slower wind below the Alfv\'en critical point, forming plasma compression and transverse flow along the field. 

Above the Alfv\'en critical point, the solar wind ram pressure overwhelms magnetic fieldline tension. The shear between faster and slower solar wind perturbs the structure in the magnetic field and the faster flow in the solar wind drags out the magnetic field, overtaking adjacent regions of slower solar wind. The fieldline connection from slower to faster wind therefore eventually creates radial inversions, or switchbacks, in the magnetic field. 

The motion of footpoints in Figure \ref{fig:f1-switchback} is driven in part by differential motion and is mediated by interchange reconnection across the \revthree{opposite edges of the} coronal hole or other regions with speed shear.  The sense of footpoint motion is largely in the direction opposite of the sense of equatorial rotation \revthree{since it is driven by differential motion}. This directionality of footpoint motion in the Super-Parker Spiral creates tangential flows in the direction of co-rotation as observed by PSP \cite[]{Kasper:2019}.

\revthree{In this picture, the orientation of tangential flows is linked to the direction of footpoint motion at the Sun
driven by differential motion, as detailed in the Introduction. The fact that coronal holes tend to rotate rigidly implies that differential motions in the photosphere drive open magnetic field footpoints into coronal holes on their leading edge (leading with respect to rigid rotation), and out of coronal holes on their trailing edge. Thus, this sense of footpoint motion consistently connects slower wind from outside the coronal hole with faster wind from within the coronal hole in the opposite sense on the leading and trailing edges of a coronal hole.  
}

\section{PSP Observations of Rarefaction Regions and Switch-backs}
\label{sec:observations}

\begin{figure}
\includegraphics[width=\columnwidth]{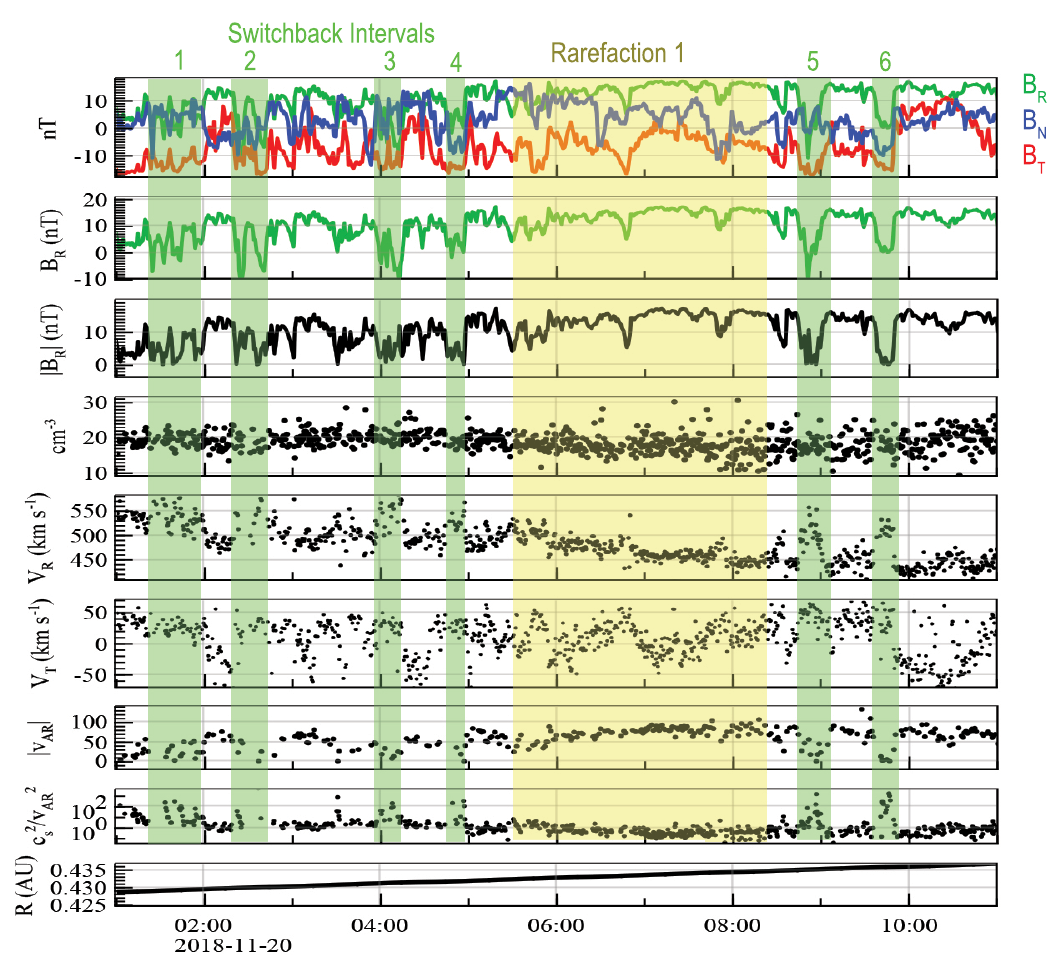} 
\caption{Observations from PSP: FIELDS data (panels 1-3), 
and SWEAP data (panels 4,5) on November 20, 2018. 
In panels 6, and 7 we form the \revfour{radial} Alfv\'en speed and \remfour{plasma beta}  
\revfour{the ratio of the sound speed to the radial Alfv\'en speed} from the SWEAP 
and FIELDS observations. Radial distance from the Sun is shown in the bottom 
panel. Green shaded regions show intervals where switchbacks are observed. 
The yellow shaded region shows a rarefaction region. 
 }
\label{fig:rarefaction1}
\end{figure}

\begin{figure}
\includegraphics[width=0.95\columnwidth]{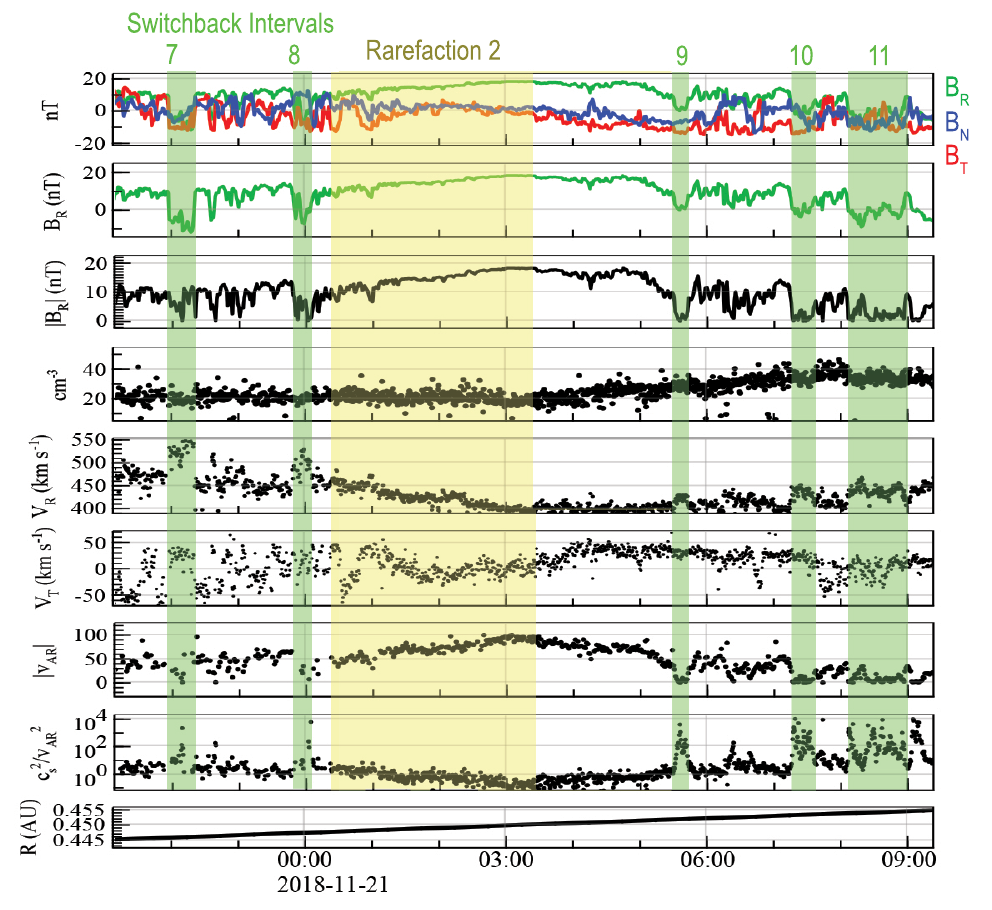} 
\caption{Observations from PSP: FIELDS data (panels 1-3), 
and SWEAP data (panels 4,5) on November 21, 2018. The format is  identical 
to Figure \ref{fig:rarefaction2}. 
 }
\label{fig:rarefaction2}
\end{figure}

We re-examine some PSP observations from Orbit 1 to test the prediction that switchbacks  occur far less often in rarefaction regions. Observations from  November 20 and November 21, 2018 are shown in Figure \ref{fig:rarefaction1} and Figure \ref{fig:rarefaction2}, respectively.   \revthree{Within these observational periods, we identify two rarefaction regions as relatively steady, multi-hour periods of decreasing radial solar wind speed. Within rarefactions it is important to separate the underlying trend of a reduction in radial wind speed from small speed fluctuations  ($<50$ km s$^{-1}$) occurring over short ($< 10$ min) periods. In a similar vein, we identify 11 switchback intervals as periods larger than $\sim$ 10 min where the radial field component vanishes or switches sign. Note that switchback intervals often include many subintervals where the radial field may switch sign or vanish intermittently.  }

 The trends within rarefactions agree with the key features associated 
with the Sub-Parker Spiral:
\begin{itemize}
\item The prediction that switchbacks should preferentially occur outside of rarefaction regions is clearly supported by the observations in Figures  \ref{fig:rarefaction1} and  \ref{fig:rarefaction2}.  The only weak possible exception is in the first rarefaction region on November 20, 2018 06:45 where we
 observe a reduction in \revfour{radial} field strength and the radial component of the magnetic
field. However, even for this event, we observe a slight increase in the radial component of the solar wind speed, suggesting that there is a Super-Parker Structure associated with the speed jump embedded within the rarefaction region.
\item Since footpoints move from the source of fast wind to slow wind in the direction opposite of co-rotation, we predict a tangential flow 
component that is negative. However, interchange reconnection and potentially changes along the flow history may lead to 
intermittent increases in the tangential flow component. Because the rarefaction region expands between faster and slower flow, 
any variations in the tangential flow will be accentuated. We therefore predicted a mix of positive and negative tangential flows, which is observed in 
both rarefaction regions. 
\item We observe an increase in the radial magnetic field  in both rarefaction regions. The second rarefaction region 
is more developed than the first, \revthree{due to the larger and longer duration drop in speed through the rarefaction, and possibly the slightly larger distance from the Sun. } In this second rarefaction, the large increase in the radial magnetic field  
relative to the other two field components is particularly evident. This strongly supports the example calculation detailed in Appendix A that the magnetic field within the Sub-Parker Spiral tends to make the  field structure more radial 
than what would be observed in a Parker Spiral.  \remfour{
There is an increase in the total magnetic field
strength observed in both rarefaction regions, which is consistent with the quantitative model results shown in the Appendix. 
} This observation  \revfour{also} directly supports the concept that the magnetic field is connected between faster and slower solar wind flow, contrary to the assumption of no footpoint motion associated with the Parker Spiral magnetic fields.
\item The increase in \revfour{radial} magnetic field strength within the Sub-Parker Spiral should increase the \revfour{radial} Alfv\'en speed\footnote{\revfour{The radial Alfv\'en speed is defined, $v_{Ar} = B_r/\sqrt{4 \pi \rho}$ where $B_r$ is the radial magnetic field, and $\rho$ is the mass density.}}
 \revfour{magnitude}
and reduce the \remfour{plasma $\beta$} \revfour{ratio of the sound speed to the radial Alfv\'en speed. } Both of these effects are observed in each rarefaction region. 
\end{itemize}

The trends  (rows 3-13 in Table \ref{table:truth}) observed where there are increases or jumps in solar wind speed
counter those within rarefactions, and agree with the Super-Parker Spiral predictions in Appendix A:
\begin{itemize}
\item Switchbacks occur  in regions where there are abrupt increases in solar wind speed, as opposed to the 
reduction in solar wind speed in rarefaction regions. 
\item Tangential flows should be positive, in the direction of co-rotation, as discussed in the previous section. 
\item The decrease in the radial magnetic field is evident in \revthree{each of the regions} identified. 
\item There is a decrease in the \remfour{total} \revfour{radial} magnetic field strength 
observed within each of \revthree{the switchback field intervals}. As shown in the Appendix, this trend supports the concept that 
the magnetic field contracts where the faster wind overtakes adjacent flow of slower solar wind. 
\item The decrease in \revfour{radial} magnetic field strength within the Super-Parker Spiral decreases the \revfour{radial} Alfv\'en speed \revfour{magnitude}
and increases the  \remfour{plasma $\beta$} \revfour{ratio of the sound speed to the radial Alfv\'en speed}, which is consistently observed in ten of the eleven switchback regions observed.
\end{itemize}

\begin{deluxetable}{cccccccccc}
\tablecaption{PSP observations within rarefactions and switchbacks \label{table:truth}}
\tablehead{
\colhead{Event} & \colhead{Fig.} & \colhead{\remfour{$|B|$}\revfour{$|B_r|$}} & \colhead{$B_r$} & \colhead{$V_r$} & \colhead{$V_t$} & \colhead{\remfour{$V_A$}\revfour{$|v_{Ar}|$}} & \colhead{\remfour{$\beta$}\revfour{$c_s^2/v_{Ar}^2$}} & \colhead{Sub-PS\tablenotemark{a} }  &   \colhead{Super-PS\tablenotemark{b} } }
\startdata
RF-1\tablenotemark{c}        & 2   &  $\uparrow$         & $\uparrow$              &    $\searrow$ &   $\uparrow$$\downarrow$     &$\uparrow$          & $\downarrow$ & 6/6 & 0/6 \\
RF-2\tablenotemark{c}        & 3   &  $\uparrow$         & $\uparrow$              &    $\searrow$ &    $\downarrow$$\uparrow$    & $\uparrow$          & $\downarrow$ & 6/6 & 0/6      \\
SB-1\tablenotemark{d}        & 2   &   $\downarrow$    & $\downarrow$         &    $\uparrow$ &   $\uparrow$       &  $\downarrow$     & $\uparrow$     &  0/6 & 6/6    \\
SB-2\tablenotemark{d}        & 2   &   $\downarrow$    & $\downarrow$         &    $\uparrow$ &    $\uparrow$       & $\downarrow$     &           ?             &  0/6 & 5/6     \\
SB-3\tablenotemark{d}        & 2   &   $\downarrow$    & $\downarrow$         &    $\uparrow$ &     $\uparrow$     & $\downarrow$     &   $\uparrow$      &  0/6 & 6/6      \\
SB-4\tablenotemark{d}        & 2   &   $\downarrow$    & $\downarrow$         &    $\uparrow$ &     $\uparrow$   & $\downarrow$     &   $\uparrow$      &  0/6 & 6/6      \\
SB-5\tablenotemark{d}        & 2   &   $\downarrow$    & $\downarrow$         &    $\uparrow$ &     $\uparrow$    & $\downarrow$     &   $\uparrow$      &  0/6 & 6/6      \\
SB-6\tablenotemark{d}        & 2   &   $\downarrow$    & $\downarrow$         &    $\uparrow$ &       $\uparrow$  & $\downarrow$     &   $\uparrow$      &  0/6 & 6/6      \\
SB-7\tablenotemark{d}        & 3   &   $\downarrow$    & $\downarrow$         &    $\uparrow$ &       $\uparrow$  &  $\downarrow$     &   $\uparrow$      &  0/6 & 6/6      \\
SB-8\tablenotemark{d}        & 3   &   $\downarrow$    & $\downarrow$         &    $\uparrow$ &       $\uparrow$   & $\downarrow$     &   $\uparrow$      &  0/6 & 6/6      \\
SB-9\tablenotemark{d}        & 3   &   $\downarrow$    & $\downarrow$         &    $\uparrow$ &        $\uparrow$  &  $\downarrow$     &   $\uparrow$      &  0/6 & 6/6      \\
SB-10\tablenotemark{d}        & 3   &   $\downarrow$    & $\downarrow$         &    $\uparrow$ &      $\uparrow$   &   $\downarrow$   &   $\uparrow$      &  0/6 & 6/6     \\
SB-11\tablenotemark{d}       & 3   &   $\downarrow$    & $\downarrow$         &    $\uparrow$ &      $\uparrow$$\downarrow$   &   $\downarrow$   &   $\uparrow$      &  1/6 & 5/6  
\enddata\tablenotetext{a}{Sub-Parker Spiral - Sub-PS}
\tablenotetext{b}{Super-Parker Spiral -- Super-PS}
\tablenotetext{c}{Rarefaction -- RF}
\tablenotetext{d}{Switchbacks -- SB}
\end{deluxetable}

PSP observations of magnetic and plasma structures show a number of common
features as summarized in Table 2. Rows RF1 and RF2 apply to rarefaction regions, and rows SB1-SB11 apply to intervals with switchbacks. 
The Sub-Parker Spiral should show enhanced \revfour{radial} magnetic field strength, a dropping radial solar wind speed, variable tangential flow, enhanced \revfour{radial} Alfv\'en speed \revfour{magnitude}, and reductions in \remfour{plasma beta}\revfour{the ratio of the sound speed to the radial Alfv\'en speed}. In contrast, the Super-Parker spiral should show the opposite trends including reduced \revfour{radial} magnetic field strength, decreased or reversed radial magnetic field component, an abrupt increase in radial solar wind speed,
positive values for the tangential flow (flows in the direction of co-rotation),
an abrupt decrease in the \revfour{radial} Alfv\'en speed \revfour{magnitude}, 
and abrupt increases in \remfour{plasma beta} \revfour{the ratio of the sound speed to the radial Alfv\'en speed}.
These predictions provide specific observational signatures, the presence of which are tested using  PSP observations. The observational signatures are related analytically and are all important for identifying the presence of the Sub-Parker Spiral or the Super-Parker Spiral. Therefore, we use these observational signatures to  ``score'' of each of the 13 intervals against all six criteria.  

The last two columns of Table 2 provide this scoring. 
For both of the rarefactions studied, each of these 6 trends are observed, and the consistency with the Sub-Parker Spiral  is six-for-six (column 8 in Table \ref{table:truth}). 
For the switchback intervals, we find perfect agreement between predicted and observed behaviors for all  but two of the intervals and even for those two switchbacks, the score was 5 out of 6. The only two exceptions include one case (switchback 2) resulting from missing or inconclusive data, and one case (switchback 11) where the tangential flows are not uniformly positive. This latter case however appears to have a more variable radial magnetic field and variable properties in general, suggesting that the interval includes an array of switchbacks embedded on smaller scales.

\section{Conclusions}
\label{sec:conk}

This study examines the development of magnetic structures in the solar wind observed by Parker Solar Probe. The Sub-Parker Spiral structure of the heliospheric magnetic field is created within rarefaction regions where footpoint motion at the Sun creates a magnetic field line connection across the gradient between fast and slow solar wind. As fast solar wind moves outward more quickly than slower solar wind, the magnetic structure across the rarefaction region becomes more radial than the Parker Spiral. We have examined  rarefaction regions observed by Parker Solar Probe and find they that are consistent with the Sub-Parker Structure. 

The direction of footpoint motion at the Sun between the source of fast and slow wind is critical in defining the magnetic structure in the heliosphere. For situations in which magnetic field footpoints move from the source of fast wind into slow wind, the magnetic structure in the heliosphere is consistent with the Sub-Parker Spiral due to the expansion within the rarefaction region and the magnetic field line connection across it. In contrast, when footpoint motion is reversed from the source of slow wind into faster wind, the magnetic structure contracts between adjacent solar wind streams and forms into a Super-Parker Spiral as the solar wind moves out into the heliosphere. The radial component of the magnetic field decreases in magnitude with distance from the Sun and eventually reverses into a switchback. 

The Sub-Parker Spiral and the Super-Parker Spiral behave functionally as opposites. The observations from Parker Solar Probe  confirm the paucity of switchbacks within  rarefaction regions. Immediately outside these rarefaction regions, we observe numerous intervals of switchbacks in the magnetic field that are directly associated with abrupt transients in solar wind speed. \revthree{In contrast to the smooth monotonic speed transition within rarefaction regions, 
the switchbacks occur in ``bursts''. The clustering and plenitude of these bursts occurring in close proximity are consistent with PSP being magnetically connected to the leading edge of the coronal hole. } The observations confirm the features of the Sub-Parker Spiral and the Super-Parker Spiral: the magnetic field strength, the radial component of the magnetic field, the speed gradients, tangential flows, Alfv\'en speed, and plasma beta all conform to predictions based on the Sub-Parker and Super-Parker Spiral within rarefaction regions and solar wind speed enhancements (spikes or jets), respectively.

Table \ref{table:truth2} describes key observations identified in the Introduction (Table \ref{table:observations}) and how the Super-Parker Spiral accounts for these observations.  In the Super Parker Spiral, the presence of footpoint motion at the Sun from the slow to fast wind source provides a configuration that leads to switchbacks as the faster wind overtakes adjacent slower wind flows. The source of faster flow can come from coronal holes, transients associated with interchange reconnection, loop sources, plumes, spicules or maco-scipules. These sources of solar wind variability have been noted as potential causes of transient jets or speed variations in the solar wind. 

The one-sided, co-rotation-directed tangential flows  observed by PSP has been one of the most important and baffling pieces of the puzzle. As shown in Figure 1, differential motion determines the orientation of tangential magnetic field variations, and the field-aligned flow below the Alfv\'en critical point. Tangential flow is thus oriented in the direction of co-rotation, naturally explaining the observed one-sided tangential flows. The field-aligned flow below the Alfv\'en critical point also causes the development of compressions, and explains the anti-correlation between density and magnetic field strength.

The Alfv\'enic correlation between velocity and field variations results from the development of Alfv\'enic structures in the solar wind. Beyond the Alfv\'en critical point, large-scale variations in the flow inevitably develop \revthree{Alfv\'enic} characteristics. Moreover, the ejection of magnetic field variations close to the Sun are consistent with Alfv\'enic structures \cite[]{Schwadron:2003}, as are the exhausts from magnetic reconnection. Therefore, whether the source involves Alfv\'en waves or interchange reconnection exhausts, an outcome is the development of Alfv\'enic variations with correlation between velocity and field deviations. 

\begin{deluxetable}{ll}
\tablecaption{PSP observations of switchbacks and transverse flows  related the Super-Parker Spiral magnetic field (mechanisms in bold 
indicate new aspects introduced in this paper). \label{table:truth2}}
\tablehead{
\colhead{Observation} & \colhead{Mechanism} }
\startdata
Switchbacked    & \textbf{Super Parker Spiral, } \\
  magnetic field            & \textbf{fast flow overtakes adjacent slower flow}  \\
                                    & \textbf{above Alfv\'en critical point} \\
                                                         \hline
  Transient jets    & Coronal holes, Interchange Reconnection\tablenotemark{a}, \\
  or pulsations      & loop sources\tablenotemark{b}, plumes\tablenotemark{c}, spicules\tablenotemark{d}  \\
 \hline
 One-sided     & \textbf{Directional  field-aligned flow  } \\
 tangential flows   & \textbf{below Alfv\'en critical point } \\
                                               \hline                                          
Transverse,  & \textbf{Differential motion drives footpoints}  \\
co-rotational flows  & \textbf{counter to co-rotation  }   \\
		\hline
Alfv\'enic correlation    &  Alfv\'enic wave development \tablenotemark{e},   \\ 
 $\delta \mathbf{v}$ to $\delta \mathbf{B}$  &  Poynting flux from Alfv\'en wave energy source\tablenotemark{f}     \\
                                               \hline
 Anti-correlated    &  \textbf{Flow compression }  \\
 $n_e$ and $|B_R|$ & \textbf{below Alfv\'en critical point} 
\enddata
\tablenotetext{a}{\cite{Fisk:2001,Fisk:2020,Zank:2020}}
\tablenotetext{b}{ \cite{Schwadron:2003,Fisk:2020,Zank:2020}}
\tablenotetext{c}{\cite{Poletto:2015}}
\tablenotetext{d}{\cite{Yamauchi:2004a} }
\tablenotetext{e}{\cite{Gosling:2009, Kasper:2019}}
\tablenotetext{f}{\cite{Schwadron:2003} }
\end{deluxetable}

Thus, this paper reveals the origin for switchbacked magnetic field structures and one-sided tangential flows from the formation of the Super-Parker Spiral, the counterpart of the Sub-Parker Spiral. These magnetic structures are produced fundamentally through footpoint motion, respectively, into and out of coronal holes at the Sun caused by differential motion and interchange reconnection across regions with a strong gradient in solar wind speed. The existence of these field structures represents a significant  departure from the standard Parker Spiral, and naturally explains fundamental relationships between their solar wind source at the Sun and the magnetic and flow structures out in the heliosphere.  

\acknowledgements
We are deeply indebted to everyone that helped make
the Parker Solar Probe (PSP) mission possible. \revthree{We thank the reviewer for their helpful comments and suggestions.}  
\revfour{We thank Dr. Stuart Bale for spotting an error in our initial plotting of the magnetic field}. 
This work was supported as a part of the PSP mission under
contract NNN06AA01C. Parker Solar Probe was designed, built,
and is now operated by the Johns Hopkins Applied Physics Laboratory as
part of NASA's Living with a Star (LWS) program (contract
NNN06AA01C). Support from the LWS management and technical team has
played a critical role in the success of the Parker Solar Probe
mission. 

\bibliographystyle{apj}

\appendix

\section{Sub-Parker and Super-Parker Spirals}
\label{sec:subparker}

\begin{figure}
\includegraphics[width=0.9\columnwidth]{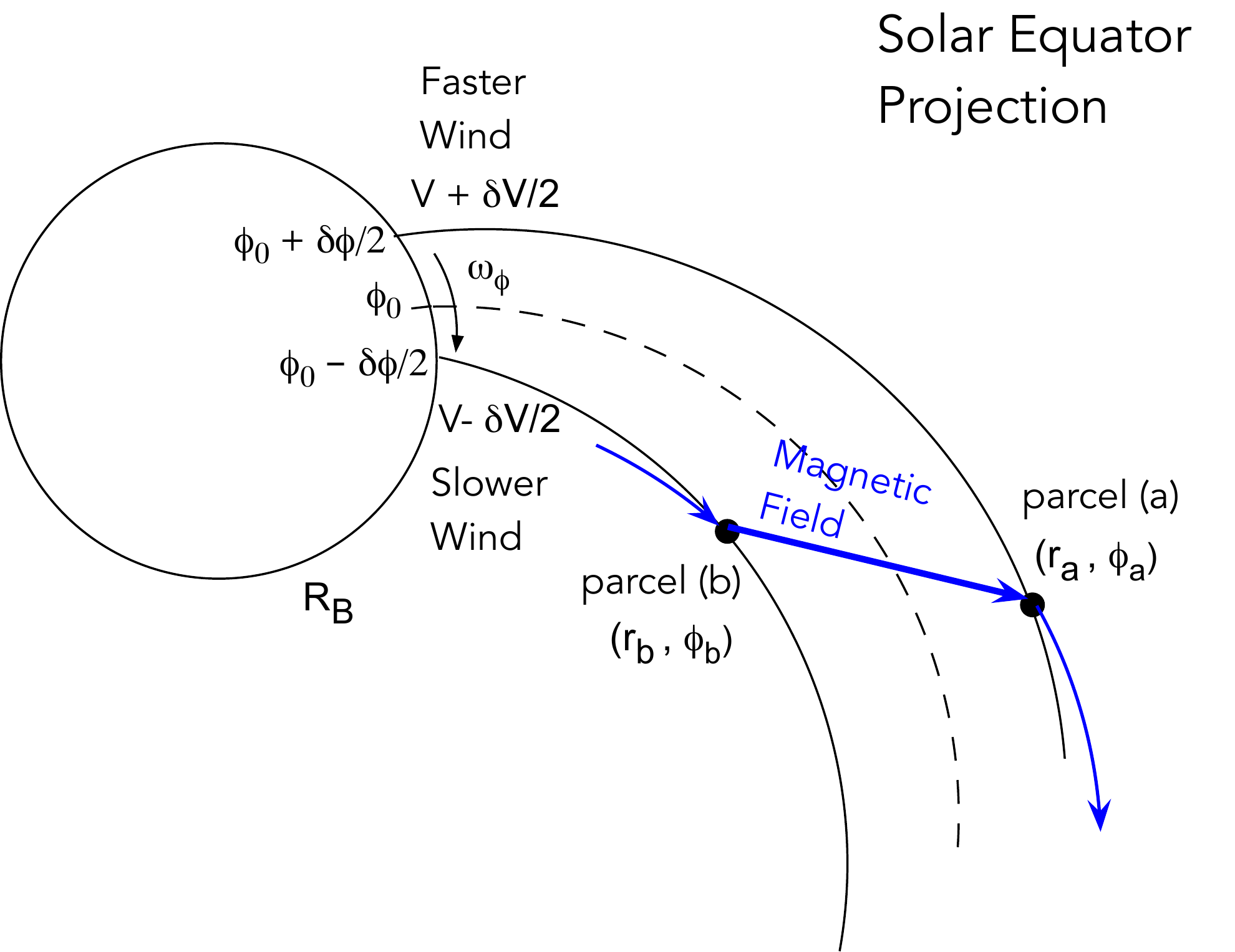} 
\caption{Footpoint motion across gradients in radial solar wind speed from faster into slower create  the conditions for  the sub-Parker
  spiral. Black-curves show the streamlines associated with parcel (a)
  and parcel (b) in the co-rotating reference frame. Footpoint motion
  provides a magnetic connection between parcel (a) and (b), which
  implies that the magnetic field is directed along the displacement
  between these plasma parcels.  This projection is on the Sun's equatorial plane.  }
\label{fig:a1}
\end{figure}

\revthree{In this appendix, we develop an analytical derivation that provides solutions for
 the Sub-Parker and the Super-Parker Spiral. }
 \cite{Schwadron:2020} follow the ballistic propagation of plasma parcels from the Sun
within the solar wind to determine the structure of the magnetic field
in the inner heliosphere.   Figure \ref{fig:a1} shows the configuration near the
Sun in the co-rotating reference frame along a boundary surface (at
radius $R_B$) where footpoint motion moves magnetic footpoints between
regions of faster wind (with speed $V+\delta V/2$) and regions of
slower wind (with speed $V - \delta V/2$). The boundary surface is at
a radius where the field has expanded and roughly reaches a uniform magnetic pressure. Footpoints rotate in the azimuthal direction at rate
$\omega_\phi = - |\omega_\phi|$ in the opposite direction of the Sun's
rigid rotation (the rigid rotation rate is $\Omega_\odot$). Footpoints
also move in co-latitude at rate $\omega_\theta = |\omega_\theta|$. 

We take the interface between fast and the
rarefaction region tilted by angle $\Psi$ with respect to the
azimuthal direction. On the inner boundary surface, at radius $R_B$,
the unit vector along the stream interface is defined
\begin{eqnarray}
  \ehat_I =  \sin \Psi \ehat_\theta + \cos \Psi \ehat_\phi. 
\end{eqnarray}
On the inner boundary the unit vector normal to the stream interface is defined, 
\begin{eqnarray}
  \ehat_\perp|_{r = R_B} =  -\cos \Psi \ehat_\theta  + \sin \Psi \ehat_\phi .
\end{eqnarray}
Therefore the footpoint rotation rate \revthree{normal to the stream interface} is
\begin{eqnarray}
\omega_\perp = - \omega_\theta \cos \Psi + \omega_\phi \sin \theta \sin\Psi . 
\end{eqnarray}
The footpoint rotation \revthree{rate times the velocity} gradient is
\begin{eqnarray}
\alpha = 
  R_B  \vec{\omega}_B \cdot \nabla V|_{r = R_B} & = & \omega_\theta \frac{\partial V}{\partial \theta} + \omega_\phi \frac{\partial V}{\partial \phi} \\
                                                  & = & R_B \omega_\perp \frac{\partial V}{\partial s_\perp}
\end{eqnarray}
where the velocity gradient normal to the interface is
\begin{eqnarray}
  R_B \frac{\partial V}{\partial s_\perp}  = \left(\omega_\theta \frac{\partial V}{\partial \theta} + \omega_\phi \frac{\partial V}{\partial \phi} \right)
           (\omega_\perp)^{-1} .
\end{eqnarray}
In the application to the rarefaction region considered here
$\omega_\phi$ is a negative quantity since longitudinal footpoint
motion opposes solar rotation, and $\omega_\theta$ is positive. This
implies that $\omega_\perp < 0 $,
\begin{eqnarray}
\omega_\perp = - ( \omega_\theta \cos \Psi  + |\omega_\phi| \sin \theta \sin\Psi) . 
\end{eqnarray}
The velocity gradient in co-latitude is a negative quantity, $\partial V/\partial
\theta < 0$, and the \revthree{velocity gradient in the azimuthal direction} is a positive quantity,
$\partial V/\partial \phi > 0$. Therefore, the velocity gradient
normal to the interface is a positive quantity,
\begin{eqnarray}
 R_B \frac{\partial V}{\partial s_\perp} = \left(\omega_\theta \left| \frac{\partial V}{\partial \theta} \right| + |\omega_\phi| \frac{\partial V}{\partial \phi} \right)
           | \omega_\perp |^{-1}  > 0 .
\end{eqnarray}
Given these properties of the velocity and footpoint rotation rates, \cite{Schwadron:2020}
expressed the magnetic field in the rarefaction region as follows
\begin{eqnarray}
  \mathbf{B} & = &  A(\mathbf{r}) \left\{
  \left( 1 + |\alpha| \frac{(r - R_B )}{V^2} \right)\ehat_r  
  -  \frac{\vec{\omega}_B r}{V} - \frac{\Omega_\Sun  r \sin \theta}{V} \ehat_\phi  \right\}
  \label{eq:B1}
\end{eqnarray}
where
\begin{eqnarray}
\vec{\omega}_B & = & \omega_\theta \ehat_\theta + \omega_\phi \sin\theta \ehat_\phi,
\end{eqnarray}
\begin{eqnarray}
A(r) = B_{fB} \left( \frac{R_B}{r} \right)^2 \left( 1 +  \frac{|\alpha|}{|\omega_\perp|}
 \frac{(r - R_B )}{V^2} \Omega_\sun  \sin \theta \sin \Psi \right)^{-1}
  \end{eqnarray}
and $B_{fB}$ normalizes the magnetic field in the ambient fast solar wind. The corresponding magnetic field in the ambient fast solar wind is:
\begin{eqnarray}
\mathbf{B}_f = B_{fB} \left( \frac{R_B}{r}\right)^2
\left( \ehat_r - \frac{\vec{\omega}_B r}{V} - \frac{\Omega_\sun r \sin\theta}{V} \ehat_\phi \right).
\end{eqnarray}

\begin{figure}
\includegraphics[width=0.9\columnwidth]{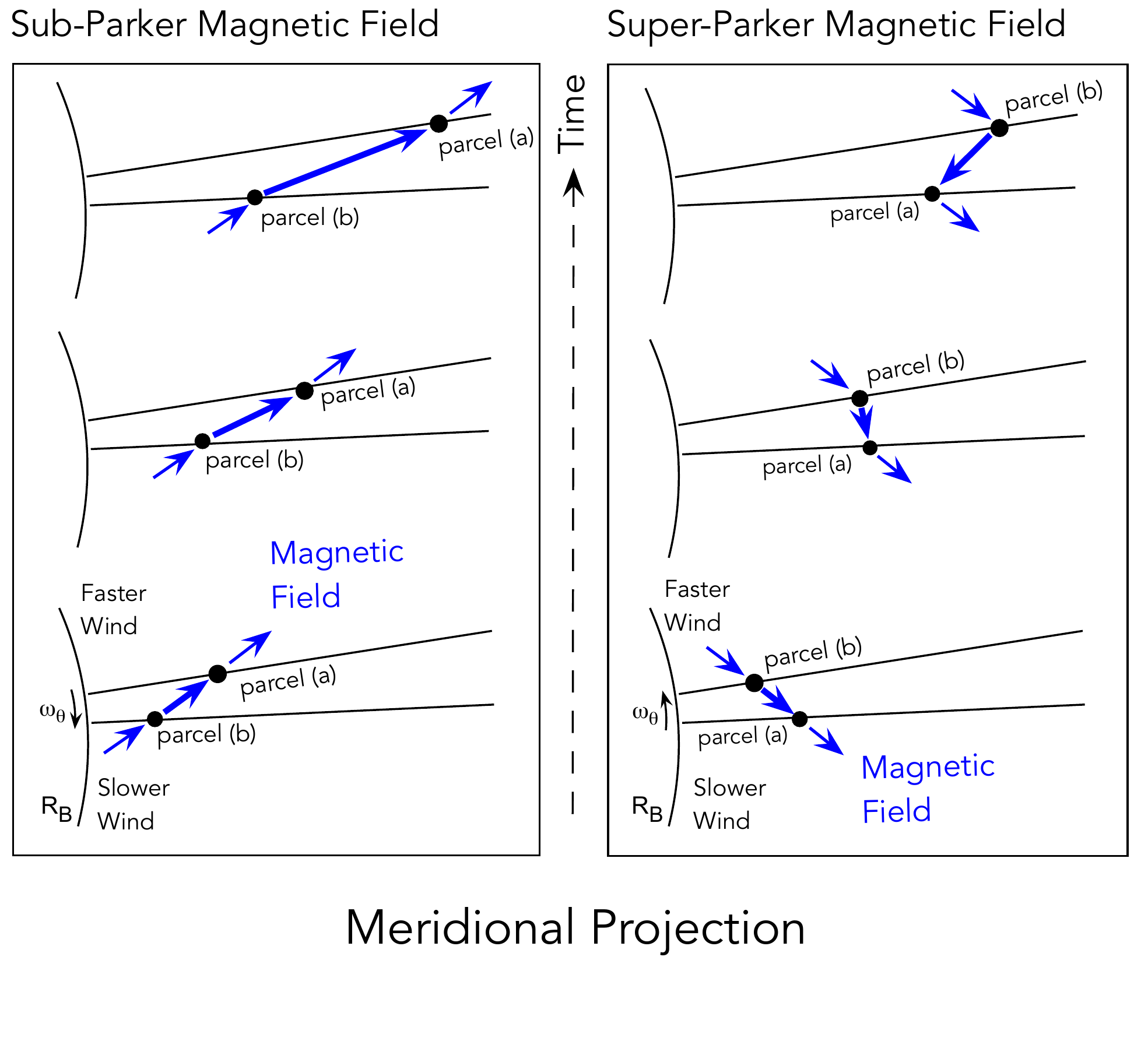} 
\caption{
\revone{
Comparison between the effects of footpoint motion from faster into slower solar wind in the case of the Sub-Parker magnetic field (left panel)  and from slower into faster solar wind in the case of the Super-Parker magnetic field (right panel). 
  Black-curves show the streamlines associated with parcel (a) released earlier and 
 parcel (b) released later after footpoints have moved in latitude in the direction indicated.  In the case (left panel) that footpoints move from 
 the source of faster solar wind  to slower wind, magnetic field lines are stretched in the radial direction forming the Sub-Parker Spiral. Note that a meridional projection is used in this figure, as opposed to the equatorial projection in Figure \ref{fig:a1}. Reversing the sense of the footpoint motion (right panel) leads to contraction of the magnetic field by the solar wind speed gradient, and ultimately leads to reversal of the field polarity associated with switchbacks. }
 }
\label{fig:a2}
\end{figure}

A specific realization of the Sub-Parker Spiral with $\Psi = 0$ reveals the 
essence of the magnetic structure. In this case, the vector perpendicular to the stream
interface is $\ehat_\perp = -\ehat_\theta$ and the Sub-Parker magnetic field is given by 
\begin{eqnarray}
  \mathbf{B}_a = B_{fB} \left( \frac{R_B}{r} \right)^2 \left\{ \left( 1 + \omega_\theta \left| \frac{\partial V}{\partial \theta} \right|_{r=R_B} \frac{(r - R_B )}{V^2} \right)\ehat_r  
  -  \frac{\vec{\omega}_B r}{V} - \frac{\Omega_\Sun  r \sin \theta}{V} \ehat_\phi  \right\}
\label{eq:Bsuper1}
\end{eqnarray}
In this expression, for simplicity we have taken the solar wind speed as a function of latitude, with higher speed wind at higher latitudes and slower solar wind at lower latitudes. The expression in equation (\ref{eq:Bsuper1}) can be generalized for any solar wind speed gradient, and any value of $\omega_\theta$:
\begin{eqnarray}
  \mathbf{B}_a = B_{fB} \left( \frac{R_B}{r} \right)^2 \left\{ \left( 1 - \omega_\theta  \left.\frac{\partial V}{\partial \theta} \right|_{r=R_B} \frac{(r - R_B )}{V^2} \right) \ehat_r  
  -  \frac{\vec{\omega}_B r}{V} - \frac{\Omega_\Sun  r \sin \theta }{V} \ehat_\phi  \right\}.  
\label{eq:Bsuper2}
\end{eqnarray}
In the case that $\omega_\theta  \left. \partial V/\partial \theta \right|_{r=R_B} < 0$, the expression conforms to the Sub-Parker Spiral solution. However, for $\omega_\theta  \left. \partial V/\partial \theta \right|_{r=R_B} > 0$, we arrive at a solution where the radial component of the magnetic field decreases with distance, and then reverses sign, leading to the formation of a switchbacked magnetic field. This is the solution for Super-Parker Spiral. With faster wind at higher latitudes, and slower wind at lower latitudes, the gradient $\left. \partial V/\partial \theta \right|_{r=R_B} < 0$. As a result, the sign of $\omega_\theta$ determines whether the solution conforms to Sub-Parker Spiral ($\omega_\theta > 0$) or a Super-Parker Spiral ($\omega_\theta < 0$). Stated more generally:  the Sub-Parker Spiral results when footpoints move from fast wind into slow wind ($\omega_\theta \left. \partial V/\partial \theta \right|_{r=R_B} < 0$); the Super-Parker spiral results when footpoints move from slow wind into fast wind ($\omega_\theta \left. \partial V/\partial \theta \right|_{r=R_B} > 0$).  Figure \ref{fig:a2} shows a meridional projection for streamlines and the resulting magnetic field configuration in the case of a Sub-Parker (left panel) and a Super-Parker (right panel). 

It is interesting to note that equation (\ref{eq:Bsuper2}), which admits both Sub-Parker and Super-Parker solutions, requires a configuration where the solar wind speed is a function of latitude, as consistent with a latitudinal shear in the solar wind. In contrast, if speed variations exist such that faster wind trails slower wind in longitude, these structures must lead to the formation of compression regions, which complicate, or disallow steady-state solutions for the magnetic field structure. The treatment of latitudinal shear in the solar wind is similar and allows straightforward insight into the structural evolution of the magnetic field.  

In the case of the Super-Parker Spiral, the switchbacked magnetic field 
appears at radial distances where
\begin{eqnarray}  
1 - |\omega_\theta | \left| \frac{\partial V}{\partial \theta} \right|_{r=R_B} \frac{(r - R_B )}{V^2} < 0 ,
\end{eqnarray}
or equivalently where 
\begin{eqnarray}
r > R_B + \frac{V^2}{ |\omega_\theta | |\partial V/\partial \theta |_{r=R_B} }.
\end{eqnarray}

As a specific example, if we consider an azimuthal motion of footpoints in the direction of solar rotation at 17\% of the solar rotation rate, a radial distance of 0.3 au, a boundary surface at $R_B = 25$ solar radii, a speed change of 100 km s$^{-1}$, and an average wind speed of 300 km s$^{-1}$, then a switchback is created if the speed gradient exists over a displacement of $< 0.8^\circ$.

\begin{figure}
\includegraphics[width=0.9\columnwidth]{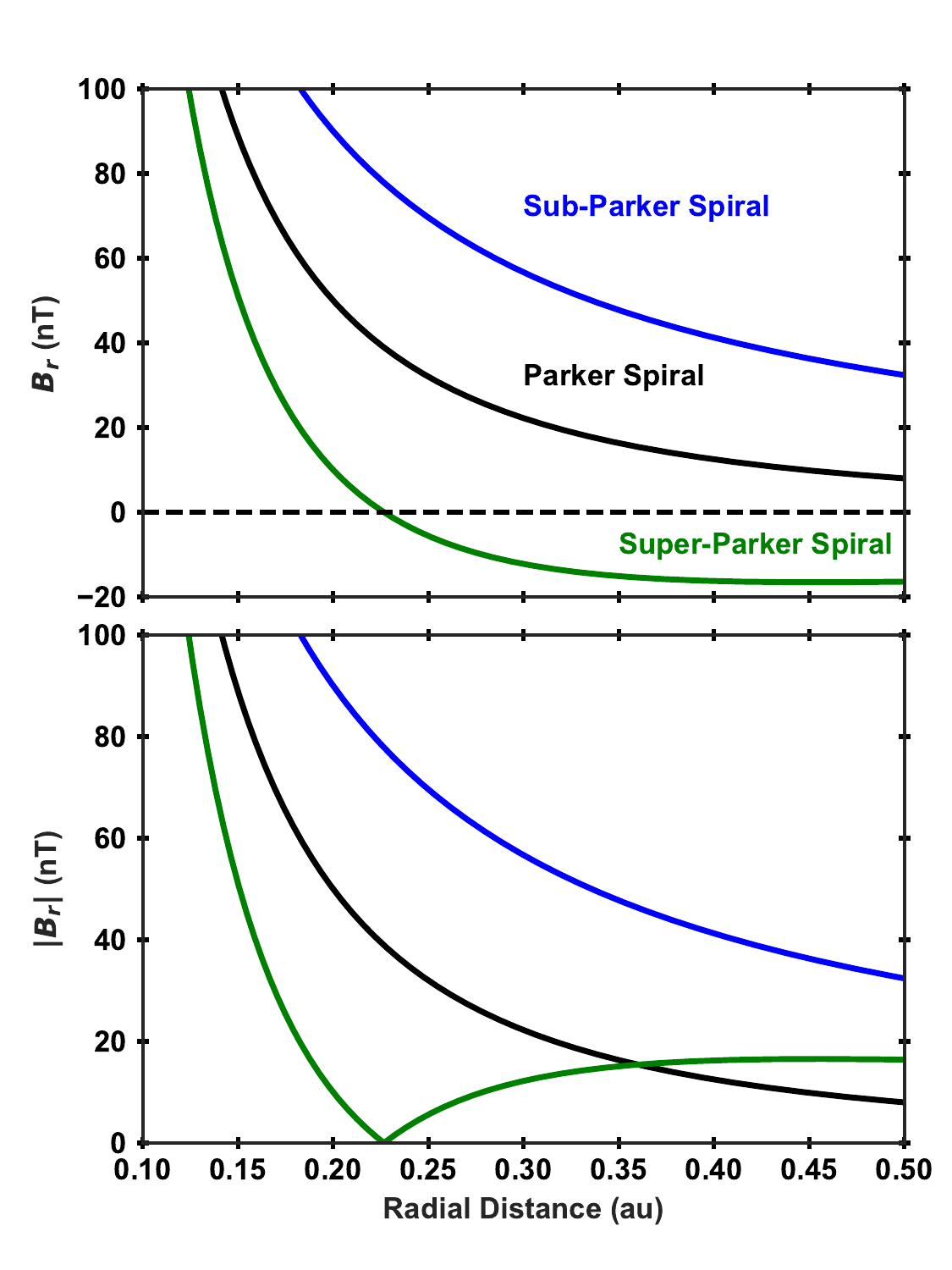} 
\caption{We use equation (\ref{eq:Bsuper2}) to solve for the evolution of the magnetic field as a function of distance from the Sun for a Parker Spiral with magnetic footpoints fixed in the co-rotating frame, a Sub-Parker Spiral with magnetic footpoints moving near the Sun from the source region of faster solar wind into the source of slower solar wind, and the Super-Parker Spiral with magnetic footpoints moving in the opposite direction from slower solar wind flow into faster solar wind flow. 
}
\label{fig:f3}
\end{figure}

In Figure \ref{fig:f3}, we show the radial component of the magnetic field and its strength as a function of radial distance based on equation (\ref{eq:Bsuper2}).  For simplicity, we consider a specific instance where there is a speed change of 50 km s$^{-1}$ across a region of 0.1$^\circ$ and an average wind speed of 400 km s$^{-1}$. Note that a 0.1$^\circ$ structure co-rotates past the a stationary observer  in $\sim 10$ minutes. The three cases shown are for the Sub-Parker Spiral (blue curve), the Parker Spiral (black curve) and the Super-Parker Spiral (green). The only factor differentiating between these instances is the rate of footpoint motion. In the case of the Parker Spiral, the footpoints are fixed and the rotation rate of footpoints is zero. For the Sub-Parker spiral the footpoints rotate from the faster wind at higher latitudes into the slower wind at lower latitudes and the rate of footpoint motion is 10\% of the solar rotation rate. For the Super-Parker Spiral, the footpoint motion is reversed, with footpoints moving from slower wind at lower latitudes into faster wind at higher latitudes, again at 10\% of the  solar rotation rate. In this case, the Super-Parker Spiral magnetic field reverses its radial component forming a switchback at 0.22 au. 

The key features associated with the three field configurations are as follows:
\begin{itemize}
\item the Parker Spiral magnetic field has a \remfour{magnitude and} 
\revfour{radial} component that drops as $\sim 1/r^2$ close to the Sun. 
\item the Sub-Parker Spiral magnetic field has a radial component \remfour{and a magnitude} that also drop as $\sim 1/r^2$ close to the Sun, but the magnitude of the \remfour{magnetic field and its} radial component \remfour{are}\revfour{is} significantly larger than that of the Parker Spiral. This increase in magnetic field strength is counter-intuitive and results from the amplification of the radial component of the magnetic field as the field is stretched radially in between the faster and slower solar wind flow. 
\item the Sub-Parker Spiral magnetic field has a radial component \remfour{and a magnitude} that also drop\revfour{s} as $\sim 1/r^2$ close to the Sun, but the \remfour{magnitude of the magnetic field and its} radial component 
\remfour{are}\revfour{is} significantly larger than that of the Parker Spiral. This increase in \revfour{radial} magnetic field strength is counter-intuitive and results from the amplification of the radial component of the magnetic field as the field is stretched radially in between the faster and slower solar wind flow. 
\item the Super-Parker Spiral initially has the weakest \remfour{field strength and the weakest} radial component of the three magnetic configurations. The radial component of the field is reduced faster than the Parker Spiral magnetic field, and ultimately reverses to form a switchback, in this case at $\sim 0.22$ au. Beyond the switchback, the radial component of the field continues to drop with distance, and the magnitude of the \revfour{radial} magnetic field ultimately overtakes that of the Parker Spiral magnetic field. This interesting increase in the \revfour{radial} field magnitude is also the result of the stretching of the reversed magnetic field between the faster and slower solar wind flow. 
\end{itemize}

It is important to note that the Sub-Parker Spiral magnetic field can persist within the rarefaction region many au from the Sun, growing to fill an increasing volume of the inner heliosphere. In contrast, the Super-Parker Spiral is associated with compression between faster and slower solar wind flow. In the case considered, where the gradient in wind speed is across latitude, the faster and slower flows can continue to stretch the field. However, this idealization is unlikely to apply over broad regions, and eventually a compression region will form that slows the faster wind and speeds up the slower wind. This compression region will subsume the Super-Parker Spiral within the inherently turbulent flow within compression regions. 

Another important feature associated with Sub-Parker Spiral and the Super-Parker Spiral is the specific alignment of flow deviations with the direction of the magnetic field. Generally, the non-radial magnetic field components will be directed opposite from the direction of footpoint motion. In the case considered with faster wind at higher latitude and slower wind at lower latitude, the non-radial field component will be directed with a positive latitudinal deviation for the Sub-Parker spiral, and a negative latitudinal deviation for the Super-Parker Spiral. However, the faster flow will lead to compression of the plasma on the flux tube beneath the Alfv\'en surface, causing a density enhancement and non-radial flow deviation in the direction of the magnetic field. The Super-Parker Spiral will therefore have non-radial flow deviations that align with the magnetic field.  In this context, the fact that PSP observes flows significantly larger  than those in the Weber–Davis model \cite[]{Kasper:2019}, and that the flow deviations are directed parallel to the magnetic field deviations is consistent with the Super-Parker Spiral. The compression side of fast-to-slow wind interface has slower wind leading faster wind in longitude, and the flow deviation is directed positively in longitude, in the same sense as co-rotation. As a result, near the source of the Super-Parker Spiral, we expect what would appear to be co-rotational flows to distances beyond the Alfv\'en critical point. 

The Sub-Parker Spiral has faster wind leading slower wind. This will tend to reduce flow deviations and form rarefaction regions as these structures evolve with distance. These weaker flow deviations and reduced solar wind density within the rarefaction region provide distinct observational features that can differentiate regions that form the Super-Parker Spiral from those that form the Sub-Parker Spiral magnetic field. 


\end{document}